# On Slice Isolation Options in the Transport Network and Associated Feasibility Indicators

Luis M. Contreras[†], Jose Ordonez-Lucena[†]
[†]Telefónica I+D / CTIO, Madrid, Spain
{luismiguel.contrerasmurillo@telefonica.com, joseantonio.ordonezlucena@telefonica.com}

*Abstract*—Isolation is one of the more relevant attributes associated to the idea of network slicing, introduced by 5G services. Through isolation it is expected that slices from different customers could gracefully coexist without interfering each other, in the sense that whatever misbehavior or unforeseen demand from one slice customer could not affect the communication service received by any other slice customer supported atop the same physical transport infrastructure. This paper surveys and compare different technical approaches that can be taken for providing distinct isolation levels in the transport network, as a major component of end-to-end network slices. Furthermore, a number of isolation feasibility indicators are defined and proposed. These indicators are based on the approaches referred before, as a mean of guiding orchestration decisions at the time of provisioning or reconfiguring the transport slices in the network.

## I. INTRODUCTION

The introduction of 5G services has motivated a series of different advances in the overall telecommunication ecosystem, ranging from novel service offerings up to innovative forms of operating the existing network infrastructure. As part of these novelties, the concept of network slicing [1] will change the way in which operators provide communication services to external customers, such as vertical industries, but also for internal business units.

With network slicing, the operators will be able to provide extreme degrees of service differentiation, by the allocation of resources and capabilities being perceived by the customers as separated and dedicated networks, in terms of capacity, resources, availability, and even control and manageability. To achieve this potential in production networks, typically built upon common, multi-service infrastructures, isolation becomes a key requirement.

Network slice isolation can be defined as the ability of a network operator to ensure that congestion, attacks and lifecycle-related events (e.g. scaling out) on one network slice does not negatively impact other existing slices. It represents a multi-faceted problem, with multiple dimensions that need to be carefully addressed, including *performance, security* and *management* [1].

According to 3GPP [2], a network slice shall be defined end-to-end, spanning across the different network domains including Radio Access Network (RAN), Core Network (CN) and Transport Network (TN). The provisioning and run-time supervision of individual slices require the definition of an overarching 3GPP management system [3], which hosts end-to-end slice orchestration functionality and interacts with the different domain-specific controllers. Each controller is responsible for the management of the different slice subnets from a given network domain. Based on the isolation requirements of these slice subnets, the domain-specific controller shall provision the corresponding Network Slice Subnet Instances (NSSIs) accordingly, allocating segregated resources in such a way these requirements are fulfilled, as depicted in Fig. 1.

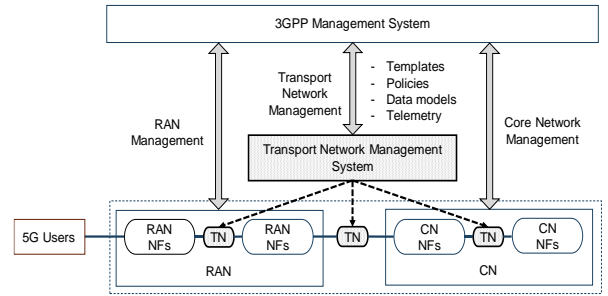

Fig. 1. 3GPP slice management and control

In this work, we focus on how the isolation problem in network slice subnets can be translated from the perspective of the TN domain, commonly referred as *transport slices*. In [4] a transport slice is defined as *"a logical network topology connecting a number of endpoints using a set of shared or dedicated network resources that are used to satisfy specific Service Level Objectives (SLOs)"*. This implies that in the particular case of the transport network isolation applies to the connectivity established between end-points where some external functions from those other constituent parts of the end-to-end network slice could be connected.

For the realization of the network slice in transport, it is expected that a Transport Network Slice Controller (T-NSC) as in [4] receives in its North Bound Interface (NBI) an abstract, technology agnostic slice request. Such request is mapped by the T-NSC to the particular technologies and topologies below, according to the specificities of the actual physical TN. However, there is not yet clearly stated what kinds of isolation can be provided by the TN, neither from the perspective of the "customer" (the overarching 3GPP management system) nor from the perspective of the provider (the system managing and controlling the TN).

As main contribution of the paper, a number of potential isolation approaches in transport network are identified, described and qualitatively compared. In addition to that, some isolation feasibility indicators are proposed to assist on orchestration decisions that require evaluating the isolation needs expressed in the transport slice requests. These indicators can help to prioritize what slices be instantiated or reconfigured first.

The reminder of this paper is as follows. Section II provides an overview of the possible isolation approaches in the transport network. In Section III, different isolation feasibility indicators are proposed as a mean for comparing transport slices. Finally, Section IV summarizes some concluding remarks and describes future lines of work.

## II. ISOLATION APPROACHES IN THE TRANSPORT NETWORK

The transport network is rich on technology options and functional capabilities. Not all the technologies offer the same level of isolation possibilities, so there are a number of alternatives to consider at the time of providing isolation. This implies that certain logic will be required at the T-NSC for decision of the most convenient option for each slice request.

The allocation of resources for achieving isolation does not imply that the allocated resources are permanently assigned (during slice lifetime) to a specific customer. That is, the concrete resource items can change along the time while the overall amount and type of resources is provided. This flexibility is maintained by the network programmability capabilities present in the network.

### A. Control plane isolation

Not all the customers will require control capabilities for the allocated slice. This will be probably limited to some advanced and specialized customers needing a deeper control of the transport slice resources. When enabling the possibility of controlling and managing the allocated resources for some customers, control plane isolation permits each of those customers to act on the assigned resources.

Network softwarization can be applied at both network provider and customer levels. In the case of the network provider, it is an enabler for the support of transport slicing, while at customer level it is an advanced feature allowing to perceive the slice as a dedicated fully controllable network. Isolation in this context refers to the ability of separating control and management concerns from different customers.

A first approach for separation requires the direct implementation on the network provider's control plane elements of different virtual spaces per slice customer, allowing each customer only access to the proper virtual space. A second, more advanced, approach includes the usage by the customer of a dedicated independent control element, interacting directly with the network provider control element. Such control plane element per customer would be limited in functionality since it is restricted to the control and management of the virtual resources assigned. Finally, that control element could be either owned by the customer itself, or facilitated by network provider (as proposed in [5]). The latter will simplify the interoperability among control plane elements, since full compatibility can be managed beforehand, while the former could require some integration effort to make the different control plane elements to interoperate.

### B. Topological isolation

Usually, topological diversity is used in order to avoid affectation in the primary and backup routes of a service on the event of failure. Thus, at the time of deploying such service, smart decisions can be taken to distinguish the resources allocated for that routes (i.e., links, nodes). This is commonly done by identifying shared risk group at resource level, then enabling to compute routes compliant with the isolation requirement.

Architectural components such as the Path Computation Element (PCE) [6] can assist on the pertinent identification of separated routes. However, when moving to the network softwarization approach, in order to ensure full isolation some additional consideration should be required to ensure that virtual topologies are effectively isolated at topological level.

This same approach of topological isolation described above can be used when dealing with overlay solutions such as the conventional VPNs.

### C. Resource isolation at device level

Different levels of isolation can also be provided at device level. The partition of the device can be considered at either hardware or software (leveraging on the available operating systems capabilities). At hardware level, for instance, different ports or boards (within associated resources such as queueing ports) could be allocated for conveying the traffic of distinct slice customers.

On the other hand, at software level, it is possible to instantiate multiple logical devices acting as virtual nodes, leveraging even in several degrees of resource differentiation at hardware level, as before. This is the case, for example, of Juniper's approaches known as logical system, virtual router or node slicing [7].

It is evident that there will always be a dependency of the same hosting device that cannot be avoided, in the sense that, depending on the case, some common parts are shared among customers. For instance, in the allocation of different ports there is dependency of the supportive board, in the allocation of boards there is dependency of the chassis (e.g., switching matrix, or fans), etc. However, these dependencies do not imply that the customers are not isolated, but that customers share the same risk group at node level.

### D. Resource isolation at data plane level

The isolation at data plane level intrinsically depends on the particular characteristics of each transport technology. Strict allocation of connectivity resources is only available in certain solutions. For instance, it could be possible to allocate for a specific slice a concrete lambda in Dense Wavelength Division Multiplexing (DWDM), or a number of calendar slots in Flex Ethernet [8].

Other technologies do not allow that kind of strict resource allocation, thus some levels of contention in the usage of shared resources could be expected. This is the case of the conventional packet-switched networks. Certain mechanisms could mitigate the contention impacts, such as advanced QoS mechanisms, either in traditional [9] or programmable [10] networks. Depending of the particular needs of the customer this could be sufficiently acceptable or not.

### E. Summary of approaches

The approaches described before can be applied independently or being combined, complementing each other and interacting in a graceful way. The final decision on which approach will be used depends on how explicit is the customer request in that respect. If not specific at all, the network provider, through the T-NSC, will take the final decision, according to predefined policies or in order to ensure some of the other SLOs requested.

It is worth noting that isolation is orthogonal to the performance SLOs, but isolation can assist on meeting those performance SLOs. For example, a SLO of throughput could be enforce through isolation (e.g., as mentioned before, dedicating a lambda of enough capacity for that customer slice) or without isolation (e.g., by grooming that traffic over lambdas with sufficient capacity to carry all the client signals).

TABLE I. SUMMARY OF ISOLATION APPROACHES

| Isolation Approach | Description | Dependencies | Scalability | Complexity | Implementation |
|---|---|---|---|---|---|
| **Control Plane** | Dedicated control of the transport resources by the customer | (Logically) centralized control element of the network operator | As the control element is software based, it can be supported as long as the computing capability is increased accordingly | The central control element should separate slice contexts among customers | Instantiation of dedicated control plane entities per slice |
| **Topology** | Diversity in the routes for a given slice | Node origin where the customer is connected | Limited by the total number of nodes and links in the network | Computation of constrained paths to ensure disjointness | Need of tools for calculating diverse paths to be allocated per slice |
| **Device** | Partition of device resources (either hardware or software) | Device common parts | Limited by the hardware resources available in the device | It is required to keep awareness of resource allocation by some control element | Association of device resources per slice |
| **Data Plane** | Allocation of connectivity resources | Port(s) where the connectivity resource is associated to | Limited to the capabilities of each technology (e.g., FlexE calendar slots) | Configuration of data plane constructs (e.g. labels or extension headers) for assignment to particular slices | Association of data plane capabilities per slice |

Table I summarizes the isolation options here described providing a qualitative comparison in terms of potential dependencies from full isolation perspective, scalability limitations, complexity issues, and generic ways for implementation of each of the approaches. Those approaches are not exclusive and could be combined for the provision of a slice with isolation requirements. Finding a simple way of comparing transport slice requests when combining more than one option can be achieved through the definition of indicators as introduced next.

### III. ISOLATION FEASIBILITY INDICATORS

It could be expected the need of handling multiple transport slice requests along the time, each of them with different needs in different aspects, including isolation.

Isolation will be impacted by the scarcity of resources not only at the time of provisioning the slice but also during its lifetime, when e.g. some network events could require the reconfiguration and reallocation of resources for the slices. In all those situations it is important to understand the feasibility of guaranteeing the demanded level of isolation for respecting the original transport slice request, especially when multiple and different resources are involved on the realization of such transport slice.

Transport slices will compete on resources of diverse nature, not becoming simple to establish comparison among them for assessing isolation feasibility along their individual lifetime. Here we introduce a modelling methodology of how feasible is to achieve the isolation of a transport slice request in order to easily compare among slices for assisting on decisions like what slice to accommodate first if reconfiguration is needed.

With that purpose, we propose first an isolation feasibility vector that could take into account different isolation levels or factors based on the particular characteristics of each slice. For simplicity, those factors are related to the isolation approaches described before.

For modelling the vector, we follow some of the ideas described in [11] but with a distinct approach. There the slices for both Radio Access (RAN) and Core Networks (CN) are characterized by different properties or traits that in some cases correspond to parameters that have a numeric, measurable value while in others are simply Boolean variables. The numeric traits can be classified as *rising*, when having a higher value is better, *falling*, the opposite, or *Gaussian*, when the values can be described by a normal-like function. In order to account for the effect of all the parameters in the same manner, those parameters become normalized, finally fitting its value into the range [0, 1]. That work then defines a vector with the obtained values that can be further merged into a single and unique value as a comparable index.

In [11], the parameters or traits are considered as defining a certain level of isolation. Example of parameters considered in that analysis are the stream cipher key's length of a radio or fiber link, the operating system of the device, the average time between vulnerabilities assessments for a router device in the CN, etc. Here, differently to that work, when referring to some transport resources being allocated to a given slice for providing isolation, it is assumed that isolation per-se is guaranteed with such resources, so no way of asserting that some slice with more dedicated resources is more isolated that another one demanding less resources. In other words, both are equally isolated at the transport layer. Thus, when applying this conceptualization to the transport network, we take a different angle. Instead of considering it for defining an isolation level, here the traits are used to obtain a comparable value of how feasible is to keep the requested isolation level along the time. That is, the higher the value, the more feasible is to accommodate the transport slice request with isolation (at either provisioning or reconfiguration time).

Thus, the parameters in this analysis basically form an isolation feasibility vector, that when merged, results into an isolation feasibility index. With that index, it is possible to easily compare among slice requests for taking informed decisions, such as reallocation of resources, by T-NSCs.

### A. Control plane

A sophisticated customer could need to have control capabilities on the resources allocated for the transport slice. Such requirement can be modeled as a Boolean variable indicating the binary option of having or not having allocated a dedicated control instance.

Defining *C* as the variable for indicating the need for a dedicated control, the values for that variable are defined as *true*, in case of needing a control instance, or *false*, on the contrary.

### B. Topology

Topology diversity enables different alternative paths in a network. To guarantee isolation the paths should be disjoint, not sharing any common single element such as node or link.

The transport slice request could lead to having topological isolation, which can be realized by reserving some specific paths for those transport slices in a manner that failures of misbehaviors in other slices do not affect those requesting isolation. Being *P* the total number of alternative and equivalent paths, a transport slice could obtain different isolation guarantees depending on the number of paths $p \in P$ that can be reserved for that slice.

The higher the value of *p*, the more difficult will be for a provider to ensure isolation for such a slice during its lifetime, since a larger number of disjoint paths is needed. *P* is a linearly increasing function with the number of *p*. However, from the perspective of isolation feasibility, the topological isolation has to be assimilated to a falling trait in the sense that a transport slice with lower requirements of topology isolation would be more feasible than other with higher requirements in this respect.

In consequence, the following normalization function $f_n(x)$ is proposed

$$f_n(x) = 1 - \left[\frac{r - l}{h - l}\right] \quad (1)$$

being *r* the requested value for the transport slice, *l* the lower possible value and *h* the highest one. It should be noted that the minimum value for topological diversity is to have available at least 2 disjoint paths, thus the rage of values for topology would be [2, *p*].

### C. Device

In the case of device partition, let's assume that there is a process of allocation of ports per transport slice. The partition can apply to both physical downlink (or client) and uplink (or network) ports for a true separation of services among customers, or other solutions as can be the allocation of physical downlink ports but the allocation of some transport construct such as a lambda in a DWDM device or a calendar slot in Flex-E links. We provide in this example the view from the perspective of a single device, but the same idea can be easily extended to a situation involving a set of devices.

Here it will be assumed that the limiting factor is on the client ports. The device will have a number of client ports of different types, but the slice is considered to require ports of the same kind and bit rate. Being *D* the total number of downlink or client ports, a transport slice could demand a given number of ports $d \in D$, allocated for that slice.

Again, the higher the value of *d*, the more difficult will be to ensure isolation for such slice during its lifetime. At both provisioning time and slice reconfiguration events, it will imply to select devices with such number of ports available. Since this trait follows a falling behavior from the perspective of isolation feasibility, the same normalization function described in Eq. (1) is used. Lower values facilitate the feasibility of the transport slice. It should be noted that the minimum value in the case of ports is 1, thus the range of admissible values is [1, *d*].

### D. Data plane

There are different technology alternatives at the data plane. For illustration, we will consider a single data plane link of 100 Gbps based on Flex-E. In the case of a Flex-E link, the full capacity of it is divided in 20 different calendar slots of 5 Gbps of capacity each. Being *S* the total number of calendar slots, a transport slice could demand a number of slots $s \in S$.

If we assume that the total capacity of the transport slice is less than 100 Gbps, a vertical could request some value in the range [1, 20] of calendar slots for its traffic. This trait is also falling in the sense that the less slots are demanded, the more feasible is to keep the isolation of the requested service along the time. Because of that, Eq. (1) is also the normalization function for this parameter.

### E. Isolation feasibbility vector example

With the examples before, it is possible to build an isolation feasibility vector in the form $\{c_i, p_i, d_i, s_i\}$, with $c_i \in C$ the need for dedicated control element instance, $p_i \in P$ the number of disjoint paths, $d_i \in D$ the number of client ports, and $s_i \in S$ the number of calendar slots for the transport slice $TS_i$.

Table II summarizes the feasibility vector for two different slice requests, for comparison. The lower and higher values of the numerical traits in this example are, respectively, *P* = [2, 4], *D* = [1, 24] and *S* = [1, 20]. As can be observed, the obtained isolation feasibility vectors obtained are $TS_1$ = {*true*, 1, 0.391, 0.947} and $TS_2$ = {*true*, 0.5, 0.521, 0.894}.

TABLE II. Example of isolation feasibility vectors

| Transport Slice Request | Trait | Requested value (*r*) | Normalized value |
|---|---|---|---|
| $TS_1$ | $c_1$ | true | -- |
| | $p_1$ | 2 | 1 |
| | $d_1$ | 15 | 0,391 |
| | $s_1$ | 2 | 0,947 |
| $TS_2$ | $c_2$ | true | -- |
| | $p_2$ | 3 | 0,5 |
| | $d_2$ | 12 | 0,521 |
| | $s_2$ | 3 | 0,894 |

Reference [11] also proposes the merging of the vectors in order for obtaining single values to facilitate comparison. A merged value from the isolation feasibility vector will be referred as an isolation feasibility index in this paper. Such a merging could consider different weights per parameter. For the example here, we consider for the merging only the numerical traits assuming equal importance and contribution of all the parameters to obtain an overall isolation feasibility index. The Boolean parameters can help to group the vectors taking advantage of the binary value of the Boolean variables, later on comparing among the ones in a group through the merged numerical value.

The function $f_m(x_1, x_2, ..., x_n)$ used for merging the values of the numerical traits in the vector is as follows.

$$f_m(x_1, x_2, ..., x_n) = n \left( \sum_{i=1}^{n} \frac{1}{X_i} \right)^{-1} \quad (2)$$

Applying Eq. (2) to the previous vectors for $TS_1$ and $TS_2$, the isolation feasibility index $TS^I_i$ are, respectively $TS^I_1 = 0{,}65$ and $TS^I_2 = 0{,}595$. Since $TS^I_1 > TS^I_2$, this can be interpreted in the way that $TS_1$ has higher feasibility, in general, than $TS_2$. Then, both the vector and the index can be used to discriminate among transport slice requests assisting on decisions for the realization of the slice. If some specific dimension of the vector is critical for the operator, the comparison of the indexes can be complemented by the comparison on a specific dimension, then permitting to qualify the feasibility of a transport slice not only on general terms but also in a particular aspect. For instance, in the example above, despite $TS_1$ having overall higher feasibility, from the point of view of the number of client ports demanded, $TS_2$ has a better indicator. If this was a critical aspect, the Transport Slice Controller could take it as valuable input for configuration decisions.

## IV. CONCLUSIONS AND FURTHER WORK

Network slicing represents a step forward the present mode of operation in service provisioning, permitting the addition of advanced characteristics. One of those advanced characteristics is the capability of isolation.

One effect of isolation in transport is that the committed SLOs for a given customer can be maintained along the service time even in the case some other customers could misbehave (e.g., by injecting more traffic than initially declared). Alternatively, the customer can request effective isolation in the sense of strict dedication of some resources because of the nature of the service (e.g., mission critical emergency services).

The first case is about guaranteeing committed SLOs. Providing such kind of guarantees in a shared network infrastructure exploiting statistical multiplexing gains is not always achievable, despite the fact of the availability of mechanisms that can help to enforce such guarantees: traffic shaping, hierarchical queuing, traffic engineering, etc.

The second situation is more about resource reservation and dedication. Reasons for that are not solely related to performance, but others such as security, service continuity and service specialization. The resource reservation and dedication does not imply that the same specific and concrete resources are devoted for a customer, but that the same kind of resources are maintained for that customer throughout the service lifetime.

The idea of transport network slice is commonly associated to that of a virtual network. However, it is important to note that the SLOs of the virtual networks are dependent on the availability and functionality of the underlay network resources in use. In that sense, the allocation of dedicated resources for a specific slice can ensure that under whatever event the virtual network running on them is not affected in any manner by other virtual networks from distinct customers. This is even more evident in the case the customer requires some capability of control for the slice, which can lead to contention when resources are shared, since multiple control elements acting on the same transport resources can create inconsistencies producing service interference.

When handling multiple transport slices at the time of provisioning or in the case of network events, it is important to define mechanisms that could allow fast decisions in case of resource scarcity. Such decisions can be part of the T-NSC logic. With that purpose some indicators are proposed, such as both the isolation feasibility vector and index, based on the isolation options defined before. These indicators can be adapted to the specific isolation options that could be in place in a certain TN, depending on the available technologies.

Next steps for this work include the mapping of the isolation requirements to the specificities of different supportive data plane technologies in the wide spectrum of transport capabilities available for a telecom operator. Final decision on what kind of isolation performed at the transport level will stay on operator's side, according to the data plane technologies and control capabilities which are available. Here again the indicators defined can assist on setting the decision criteria for prioritizing orchestration actions.

ACKNOWLEDGMENT

This work has been partly funded by the 5G-PPP projects 5G-DIVE (Grant Agreement no. 859881), 5GROWTH (G. A. no. 856709) and 5G-VINNI (G. A. no. 815279).

REFERENCES

[1] J. Ordonez-Lucena, et al., "Network slicing in for 5G with SDN/NFV: Concepts, Architectures, and Challenges", *IEEE Communications Magazine*, vol. 55, no. 5, pp. 80-87, May 2017.

[2] 3GPP TS 23.501, "System architecture for the 5G System (5GS)".

[3] 3GPP TS 28.531, "Management and orchestration; Provisioning".

[4] R. Rokui, et al., "Definition of IETF Network Slices", draft-nsdt-teas-ietf-network-slice-definition-02 (work in progress), December 2020.

[5] S. Clayman, et al., "WIM on-demand – A modular approach for managing network slices", *IEEE Conference on Network Softwarization* (NetSoft), Ghent, Belgium, June 2020.

[6] A. Farrel, J.-P. Vasseur, J. Ash, "A Path Computation Element (PCE)-Based Architecture", RFC4655, August 2006.

[7] Juniper, "Logical Systems User Guide for Routers and Switches", March 2020. Available at: https://www.juniper.net/documentation/en_US/junos/information-products/pathway-pages/config-guide-logical-systems/config-guide-logical-systems.pdf

[8] OIF, "Flex Ethernet 2.0 Implementation Agreement", IA OIF-FLEXE-02.0, June 2018.

[9] J. Soldatos, E. Vayias, G. Kormentzas, "On the Building Blocks of Quality of Service in Heterogenus IP Networks", *IEEE Comm. Surveys & Tutorials*, Vol. 7, No. 1, pp. 70-89, First Quarter 2005.

[10] M. Karakus, A. Durresi, "Quality of Service (QoS) in Software Defined Networking (SDN): A survey", *Journal of Network and Computer Applications*, Vol. 80, pp. 200-218, 2017.

[11] Z. Kotulski, et al., "5G networks: Types of isolation and their parameters in RAN and CN slices", *Computer Networks*, Vol. 171, April 2020.